\documentstyle[amssym]{CUPconf}

\ifoldfss
\else
  \ifnfssone
    \newmathalphabet{\mathit}
      \addtoversion{normal}{\mathit}{cmr}{m}{it}
      \addtoversion{bold}{\mathit}{cmr}{bx}{it}
    \newmathalphabet{\mathcal}
      \addtoversion{normal}{\mathcal}{cmsy}{m}{n}
    \else
    \ifnfsstwo
    \fi
  \fi
\fi

\def\arcdeg{\hbox{$^\circ$}}
\def\arcmin{\hbox{$^\prime$}}

\def\hexnumber#1{\ifcase#1 0\or1\or2\or3\or4\or5\or6\or7\or8\or9\or
 A\or B\or C\or D\or E\or F\fi }

\font\eurmten=eurm10
\font\eurmseven=eurm10 at 7pt
\font\eurmfive=eurm10 at 5pt
\newfam\eurmfam
\textfont\eurmfam=\eurmten
\scriptfont\eurmfam=\eurmseven
\scriptscriptfont\eurmfam=\eurmfive
\edef\eurm@{\hexnumber\eurmfam}

\mathchardef\upi="0\eurm@19   
\mathchardef\umu="0\eurm@16   

\font\msxten=msam10
\font\msxseven=msam10 at 7pt
\font\msxfive=msam10 at 5pt
\newfam\msxfam
\textfont\msxfam=\msxten
\scriptfont\msxfam=\msxseven
\scriptscriptfont\msxfam=\msxfive
\edef\msx@{\hexnumber\msxfam}

\mathchardef\leqslant="3\msx@36
\mathchardef\geqslant="3\msx@3E

\def\lesssim{\mathrel{\hbox{\rlap{\hbox{\lower4pt\hbox{$\sim$}}}\hbox{$<$}}}}
\def\gtrsim{\mathrel{\hbox{\rlap{\hbox{\lower4pt\hbox{$\sim$}}}\hbox{$>$}}}}
\makeatletter
\ifx\CUP@mtlplain@loaded\undefined
\else
\fi
\makeatother
\ifnfsstwo

\fi
\ifnfssone

\fi
\ifoldfss

\fi

\mathchardef\varLambda="0103

\makeatletter
\ifx\CUP@mtlplain@loaded\undefined
\else
\fi
\makeatother
\title[Integrated Stellar Populations of Bulges]{Integrated Stellar
  Populations of Bulges: First Results}

\author[S. C. Trager et al.]{S.\ns C.\ns T\ls R\ls A\ls G\ls E\ls
R$^1$,\ns J.\ns J.\ns D\ls A\ls L\ls C\ls A\ls N\ls T\ls O\ls
N$^2$,\\\and\ns B.\ns J.\ns W\ls E\ls I\ls N\ls E\ls R$^1$}

\affiliation{$^1$Carnegie Observatories, 813 Santa Barbara Street,
Pasadena CA 91101\\[\affilskip]
$^2$Department of Astronomy, University of Washington, Box 351580,
Seattle WA 98195-1580}

\begin{document}
\ifnfssone
\else
  \ifnfsstwo
  \else
    \ifoldfss
      \let\mathcal\cal
      \let\mathrm\rm
      \let\mathsf\sf
    \fi
  \fi
\fi

\maketitle

\begin{abstract}
We present first results from an on-going survey of the stellar
populations of the bulges and inner disks of spirals at various points
along the Hubble sequence.  In particular, we are investigating the
hypotheses that bulges of early-type spirals are akin to (and may in
fact originally have been) intermediate-luminosity ellipticals while
bulges of late-type spirals are formed from dynamical instabilities in
their disks.  Absorption-line spectroscopy of the central regions of
Sa--Sd spirals is combined with stellar population models to determine
integrated mean ages and metallicities.  These ages and metallicities
are used to investigate stellar population differences both between
the bulges and inner disks of these spirals and between bulges and
ellipticals in an attempt to place observational constraints on the
formation mechanisms of spiral bulges.
\end{abstract}

\firstsection 

\section{Introduction}

Current thinking considers two major pathways to the formation of
spiral bulges.  Simplistically, either the bulge formed before the
disk (``bulge-first,'' e.g. van den Bosch 1998), or formed {\it from}
the disk (``disk-first,'' e.g., Combes \& Sanders 1981).
Previous studies have shown that bulges of big-bulge spirals (like
M31) share at least some stellar population properties with mid-sized
elliptical galaxies.  They fall along the $D_n$--$\sigma$ relation
(Dressler 1987) and the Fundamental Plane (Bender, Burstein \& Faber
1992).  Moreover, Jablonka et al.\ (1996) and Idiart et al.\ (1996)
find that bulges of spirals (as late as Sc) fall along the
Mg--$\sigma$ relation defined by early-type galaxies, suggesting that
bulges share a mass-metallicity relation with elliptical galaxies.
Together with kinematical evidence, these observations have led to the
idea that mid-sized ellipticals accrete disks from some leftover gas
reservoir, forming spiral galaxies (see, e.g., Kauffmann, White \&
Guiderdoni 1993).  Therefore, the stars in spiral galaxy bulges should
be similar to the stars in ellipticals (i.e.\ metal-rich, high
[Mg/Fe]; e.g., Worthey, Faber \& Gonz{\'a}lez 1992; Trager et al.\
1998b) and unlike the stars in spiral galaxy disks.

Kinematical and surface brightness observations of small-bulge spirals
indicate that their bulges share many properties with their disks:
Many bulges in these spirals are better represented by a shallower
exponential profile than by the steep de Vaucouleurs profile (e.g., de
Jong 1996) and many of these bulges exhibit disk-like kinematics (see
Kormendy 1993 for an excellent review).  Moreover, Peletier \&
Balcells (1996) and de Jong (1996) find that the color difference
between bulge and disk in an individual galaxy is much smaller than
the variation in colors across galaxies of a single Hubble
type.\footnote{However, very recent observations by Peletier \& Davies
(this meeting), using HST WFPC2 and NICMOS imaging of the Peletier \&
Balcells sample of early-type spirals, have shown that bulges are
uniformly red, suggesting that the bulges are uniformly old and
metal-rich.  Clearly these issues are by no means settled with current
observational data.}  These observations have led these authors and
others to propose that bulges (of small-bulge spirals at least) are
formed from the stars already present in their underlying disks.  That
is, the stars in spiral galaxy bulges should be similar to the stars
in spiral galaxy disks (e.g., ages and metallicities of their inner
disks, solar [Mg/Fe]) and unlike stars in ellipticals.

These two scenarios do have testable consequences, and sophisticated
techniques now available can provide definitive answers to these
questions.  We have embarked on a multi-year survey to probe the
stellar content of the bulges and inner disks of spirals using
absorption-line strengths.  These measurements can provide the
critical tests needed to understand the basic mechanisms driving bulge
formation.

\section{The stellar populations of spiral bulges}

\subsection{Sample and observations}

We have selected a sample of 91 southern face-on spirals, ranging in
type from S0/a to Sdm, barred and unbarred, from the ESO (B) survey.
These spirals are not interacting, have major axes larger than
3\arcmin, are within 4000 ${\rm km\;s^{-1}}$ and are located at
$|b|>20\arcdeg$.  The observations have been made using the long-slit
Boller \& Chivens spectrograph at Las Campanas Observatory in the
blue, giving $\approx2$ \AA\ ($\sigma_{\rm inst}\approx35\;{\rm
km\;s^{-1}}$) resolution in the 4000--5200 \AA\ region.  This spectral
range covers portions of both the Lick/IDS (Burstein et al.\ 1984;
Worthey et al.\ 1994; Trager et al.\ 1998a) and Rose (1985, 1994)
absorption-line strength systems.  To date, ten (primarily unbarred)
spirals have been observed, most along both major and minor axes, and
data from four have been processed.  Line-strength profiles for these
four galaxies have been calibrated onto the Lick/IDS system using
stellar observations in common with Jones (1996), and velocity
dispersion profiles and rotation curves have been derived following
the Fourier-quotient procedure described by Gonz{\'a}lez (1993).

\subsection{The Mg--$\sigma$ relation}

As described above, bulge-first formation mechanisms find support in
the observation that bulges in galaxies as late as Sc fall along the
same Mg$_2$--$\sigma_0$ relation as early-type galaxies (Jablonka et
al.\ 1996; Idiart et al.\ 1996), suggesting the existence of a
mass-metallicity relation in spirals and a narrow spread in ages at
fixed $\sigma_0$.  We confirm these observational results: our bulges
fall on the Mg$b$--$\sigma_0$ relation defined by early-type galaxies
(our observations do not cover the red sideband of the broad Mg$_2$
index; see data in Trager et al.\ 1998a).  However, the Mg--$\sigma$
relation is inherently {\it degenerate to compensating variations in
metallicity and age in old stellar populations\/}---large age spreads
can exist if there is a complementary age-metallicity relation
(Worthey, Trager \& Faber 1996; Trager 1997).  This intrinsic
age-metallicity degeneracy in Mg$b$ and Mg$_2$ (and other metal lines
and broadband colors) prevents the Mg--$\sigma$ relation from being an
effective stellar population age discriminator.  More sophisticated
techniques are obviously necessary to distinguish between bulge
formation mechanisms.


\subsection{Balmer-metal line diagrams}

\begin{figure}[t]
\begin{center}
\begin{picture}(100,200)(50,30)
\put(0,0){\includegraphics{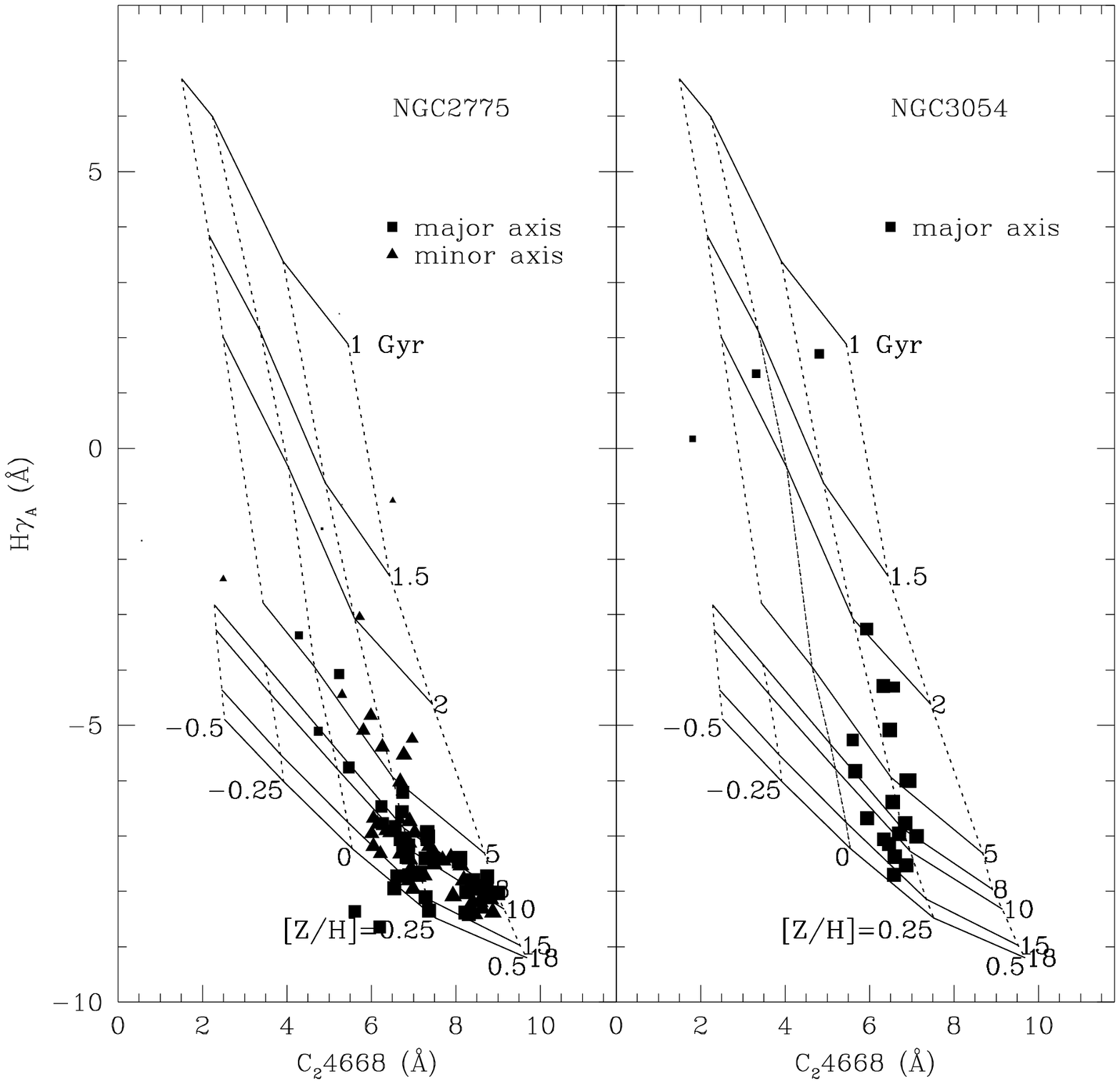} \hfil \includegraphics{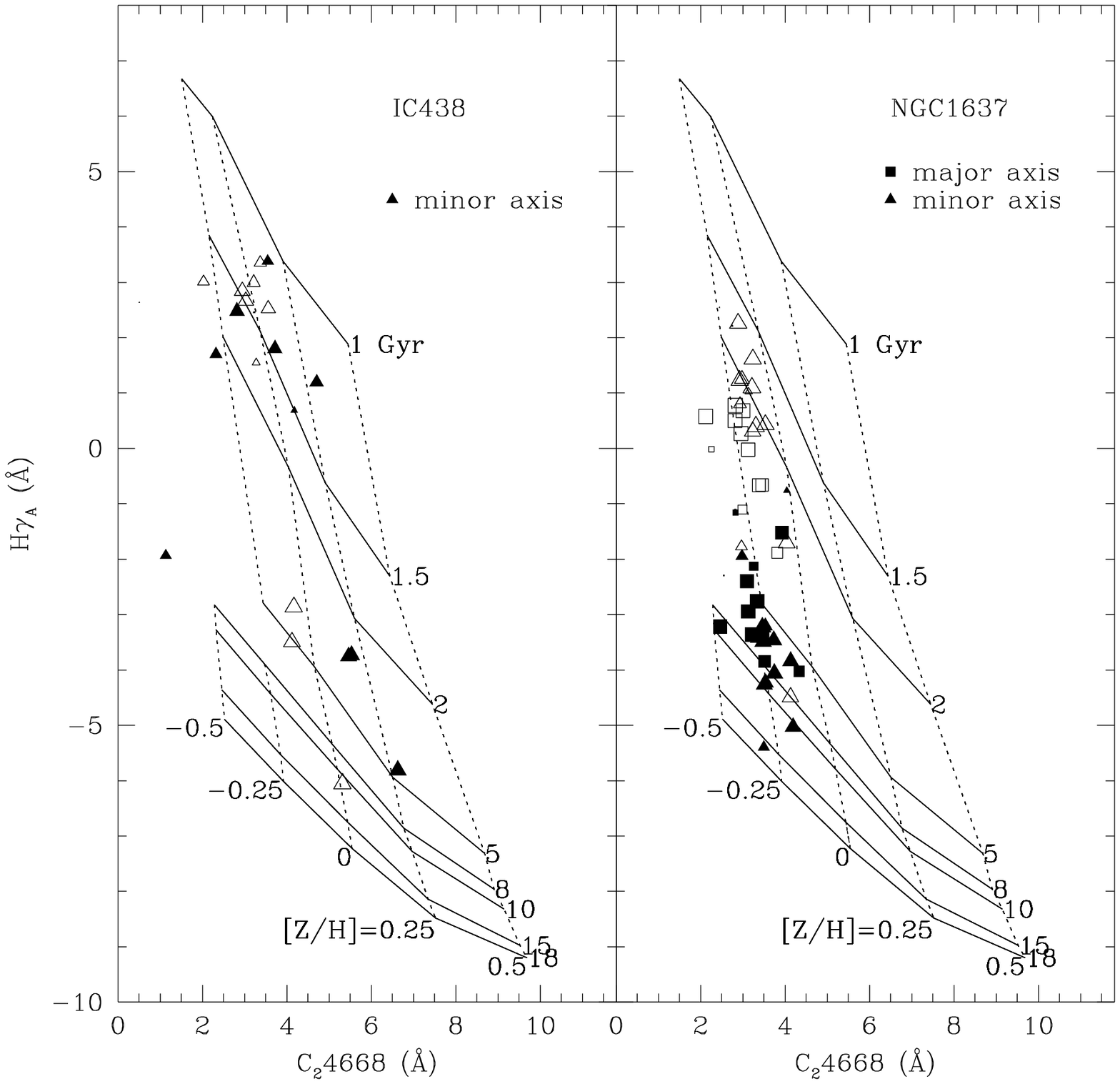}}
\noindent
\end{picture}
 \caption{Balmer--metal-line diagrams for two Sa--Sb spiral bulges
 (NGC 2775 and NGC 3054, left panels) and two Sc bulges (IC 438 and
 NGC 1637, right panels).  The Balmer line index H$\gamma_A$ is a
 sensitive measure of the presence of intermediate-age (1--10 Gyr)
 stars in an old population.  The metal-line index $C_24668$ is a
 sensitive measure of the bulk metallicity of an old population.
 Although both indices are slightly sensitive to both age and
 metallicity, used together they can effectively measure the mean age
 and metallicity in an old stellar population (Worthey 1994; Worthey
 \& Ottaviani 1997; Trager et al., 1998).  Points are coded by the
 axis along which the slit was placed, and grow smaller going out from
 the center.  Closed symbols are as observed; the open symbols are an
 attempt to correct for the emission fill-in in the H$\gamma_A$ index
 using an optimized stellar template (cf.\ Gonz{\'a}lez 1993).  Lines
 represent models of Worthey (1994) and Worthey \& Ottaviani (1997):
 solid lines are contours of constant age and dotted lines are
 contours of constant metallicity.  The Sa/Sb bulges on the left are
 clearly quite old (10--15 Gyr) and metal-rich (metallicities as high
 or higher than solar) and are older than their disks by at least a
 few Gyr.  In contrast, the Sc bulges seem younger than the Sa--Sb
 bulges ($<10$ Gyr), and the bulge of NGC 1637 is nearly as young as
 its disk.  Both Sc bulges are also slightly metal-poor (metallicities
 less than solar), suggesting that a mass-metallicity relation may
 exist for spiral bulges, as suggested by the Mg--$\sigma$ relation.}
\label{bmd}
\end{center}
\end{figure}

There is a way to break the age-metallicity degeneracy---Balmer lines
are more age-sensitive than metal-line indices (with the possible
exception of the G band; Worthey 1994).  In the absence of of nebular
emission or large numbers of blue horizontal branch stars or blue
stragglers, Balmer-line strengths reflect light-weighted mean turnoff
temperature of the composite stellar population---i.e., the mean age
of the population.  Balmer lines do have some intrinsic metal
dependence, however, so a clean separation of age and metallicity
effects requires diagnostic diagrams combining Balmer line and metal
line strengths.  Figure~\ref{bmd} presents such diagrams for the first
four galaxies analyzed in our sample.

The Balmer--metal-line diagrams for large bulges (Sab--Sbc,
$\sigma_0>100\;{\rm km\;s^{-1}}$) suggest that the bulges of these
early-type spirals are consistent with having old, metal-rich stellar
populations, as seen in our own galaxy ($t\gtrsim10$ Gyr, ${\rm
[C/H]}\gtrsim0$; e.g., McWilliam \& Rich 1994; Bruzual et al.\ 1997).
These bulges are also older than their inner disks by several Gyr,
suggesting that these bulges formed early on in the galaxies'
histories, and that bulge-first formation is a likely scenario for
these galaxies.

On the other hand, small bulges (Sc, $\sigma_0<100\;{\rm km\;s^{-1}}$)
seem to have younger and more metal-poor populations ($t\lesssim10$
Gyr, ${\rm [C/H]}\lesssim0$), and at least in some galaxies (NGC
1637), the bulge and inner disk are roughly the same age.  Such an
observation provides strong support for a common origin of bulge and
disk material---that is, for a disk-first formation scenario.
However, contamination from emission is difficult to remove,
separation of bulge and disk light has not yet been attempted, and
small amounts of young or intermediate-age populations can
significantly increase Balmer-line strengths (Trager et al.\ 1998b).
In these late-type, star-forming, small-bulged galaxies, such effects
may obscure truly old bulge populations.

\section{Conclusions}

The initial indications are that large bulges are genuinely old,
metal-rich systems, as expected from our own galactic bulge.  The
stellar populations of small bulges appear to be younger and more
metal-poor than the large bulges, but contamination from emission
lines and disk light make these conclusions uncertain for the moment.
If these results stand with more data and more extended analysis, we
may find that bulges may have more than one formation mechanism---or
even more than one mechanism may occur in a single galaxy (Rix, this
meeting).

\begin{acknowledgments}
We thank our collaborators Dr.\ R. O. Marzke and Dr.\ A. McWilliam for
many stimulating converations during the course of this work.  SCT
would like to thank the organizers for holding this very interesting
workshop and for their financial support.
\end{acknowledgments}

\end{document}